\documentclass{article}
\usepackage[english]{babel}
\usepackage[utf8x]{inputenc}
\usepackage[T1]{fontenc}
\usepackage{setspace}

\usepackage[letterpaper,top=2.5cm,bottom=2.5cm,left=3cm,right=3cm,marginparwidth=1.75cm]{geometry}

\usepackage{amsmath}
\usepackage{graphicx}
\usepackage[colorinlistoftodos]{todonotes}
\usepackage[colorlinks=true, allcolors=blue]{hyperref}

\title{Searching for $0\nu\beta\beta$ decay in $^{136}$Xe --\\ towards the tonne-scale and beyond}
\author{T. Brunner$^{1,2}$ and L. Winslow$^{3}$}
\date{%
    $^1$Physics Department, McGill University, Montreal, QC, Canada\\%
    $^2$TRIUMF, Vancouver, BC, Canada\\%
    $^3$Laboratory for Nuclear Science, Massachusetts Institute of Technology, MA, USA\\[2ex]%
    \today
}
\begin{document}
\maketitle

\begin{abstract}
The quest for neutrinoless double-beta decay ($0\nu\beta\beta$) is a promising experimental approach to search for lepton number violation in weak interactions, a key ingredient in generating the matter-antimatter asymmetry through models of Leptogenesis. The $^{136}$Xe-based $0\nu\beta\beta$ experiments KamLAND-Zen and EXO-200 currently set the most stringent limits on this process using two very different techniques. Each are preparing the next generation experiment, which will search for $0\nu\beta\beta$ in the parameter space corresponding to the inverted hierarchy for neutrino mass. Both of these techniques scale well to larger volumes while incorporating interesting new techniques. We present the status of current and next generation experiments of these collaborations and present two developments with the potential to identify $\beta\beta$ decay events.
\end{abstract}

\section{Introduction}
Neutrinos are one of the least understood particles in the universe, yet almost as abundant as photons. They only interact weakly, which makes experiments aimed at determining their properties extremely difficult. Neutrinos are electrically neutral, which makes them unique among all known fermions and offers the possibility that they may in fact be Majorana particles, i.e., neutrino and anti-neutrino could be identical particles. The only currently feasible approach to determine the Majorana nature of neutrinos is by searching for the lepton number violating decays, such as neutrinoless double-beta decay ($0\nu\beta\beta$). A positive observation of $0\nu\beta\beta$ would demonstrate that lepton number is not a conserved quantity in weak interactions and prove that the neutrino is a Majorana fermion. This new physics would provide a mechanism through Leptogenesis for generating the matter-antimatter asymmetry in the universe, answering the question of why we live in a matter-dominated universe. 

Double beta decay occurs in 35 isotopes \cite{Tretyak:2002dx}, but only a few of them are of interest for $0\nu\beta\beta$ searches due to considerations of endpoint and natural abundance (see \cite{DellOro2016} for a list of isotopes and past, current and future $\beta\beta$ decay experiments). The Enriched Xenon Observatory (EXO) and KamLAND-Zen experiments are searching for a $0\nu\beta\beta$ decay in $^{136}$Xe. Xenon-136 has a relatively high natural abundance of 8.6\%, which makes enrichment easier, and the $Q$ value of 2.5\,MeV is above most naturally occurring backgrounds. 

\begin{figure}
	\centering
		\includegraphics[width=.6\textwidth]{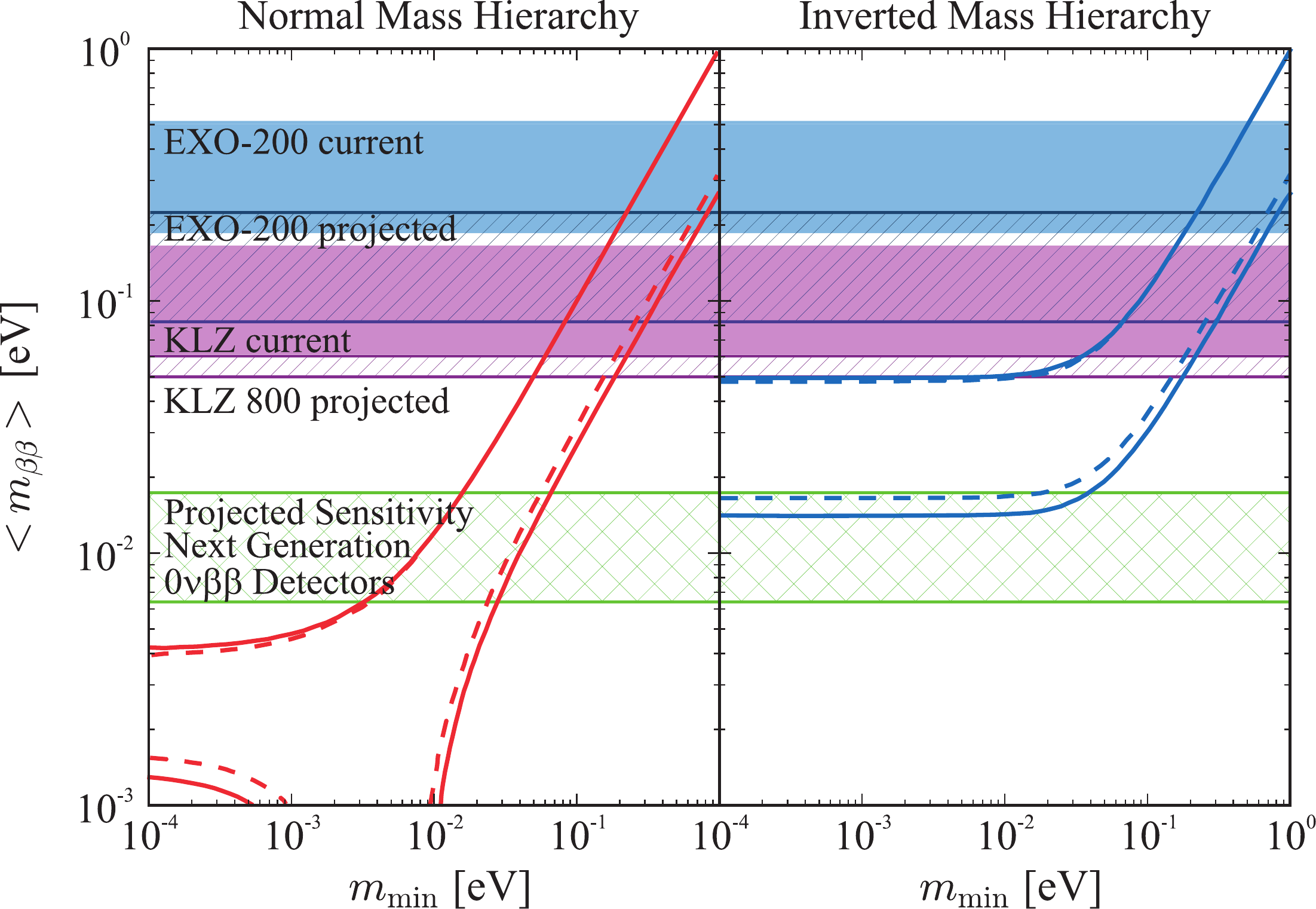}
	\caption{Measured (solid) and projected (hatched) effective Majorana neutrino mass sensitivity limits of EXO-200 and KamLAND-ZEN (KLZ) as function of the lightest neutrino mass eigenstate $m_{min}$. The sensitivity of next generation $^{136}$Xe $0\nu\beta\beta$ decay experiments is shown as cross-hatched band. The allowed parameter space from oscillation experiments is shown as red and blue band for normal and inverted mass hierarchy, respectively. 
	\label{fig:sensitivity-comp}}
\end{figure}

Once observed, the effective Majorana neutrino mass $\langle m_{\beta\beta}\rangle$ can also be extracted from the $0\nu\beta\beta$ rate
\begin{equation}
\Gamma_{1/2}^{0\nu}=G_{0\nu}\ |M^{0\nu}|^2\ |\langle m_{\beta\beta}\rangle|^2 \ ,
\end{equation}
where $G_{0\nu}$ is the phase-space factor and $M^{0\nu}$ is the nuclear matrix element. Both values are provided by nuclear theory, although with sizable differences between nuclear matrix elements calculated in different theoretical frameworks. The sensitivity of experiments is quoted in terms of $\langle m_{\beta\beta}\rangle$ as shown in Figure\,\ref{fig:sensitivity-comp} with increased experimental sensitivity corresponding to smaller $\langle m_{\beta\beta}\rangle$. 

The EXO and KamLAND-Zen collaborations are currently developing concepts of next generation experiments in parallel to the operation and data taking with the current detectors. Current and future developments will be presented in the following sections.
\section{Current Results}
The current half-life limits of EXO-200 and KamLAND-ZEN were used to extract the effective Majorana neutrino mass-limit region using nuclear matrix elements from \cite{Mustonen2013,Rodriguez2010}, which is shown as solid bands in Figure\,\ref{fig:sensitivity-comp} as a function of the lightest neutrino mass eigenstate. Projected sensitivities of EXO-200 final and KamLAND-Zen 800 are shown as hatched bands.
\subsection{EXO-200}
\begin{figure}
	\centering
		\includegraphics[width=1.\textwidth]{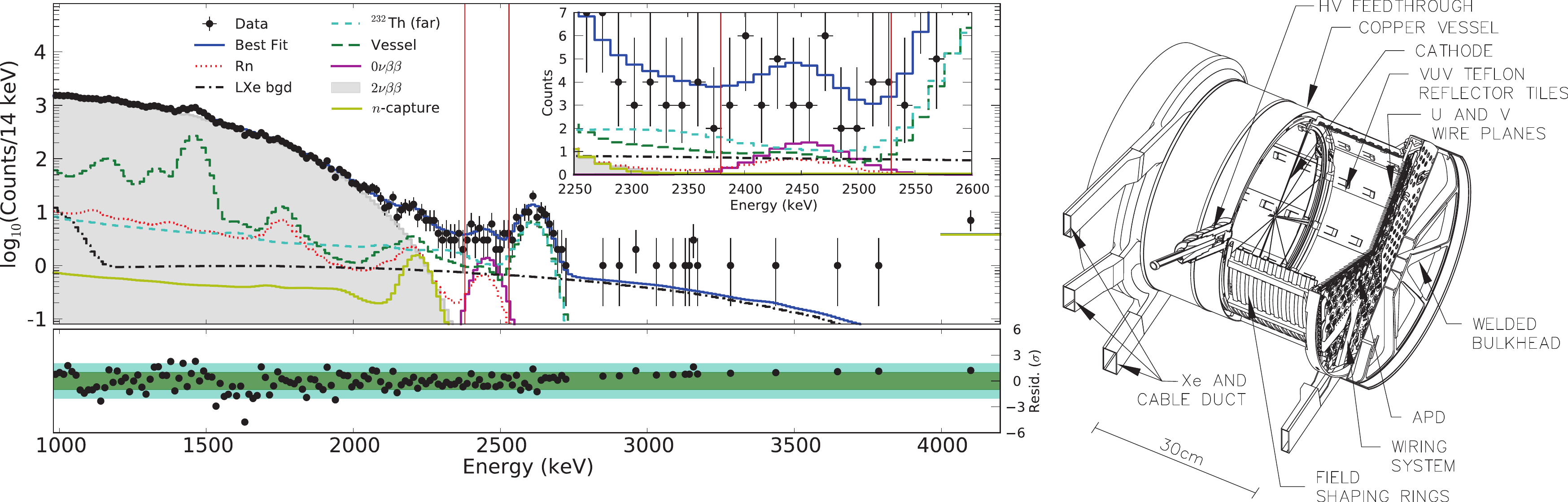}
	\caption{(left) EXO-200 single-site energy spectrum. The insert is zoomed in at the region around the $Q$ value. (right) Sectioned view of the EXO-200 TPC with annotations to main features of the detector. The Teflon sheet in front of the field-shaping rings reflects scintillation light and increases the APD's acceptance. Figures adapted from \cite{Albert2014,Albert2014PRC}.}
	\label{fig:EXO-200}
\end{figure}
EXO-200 is a liquid xenon time-projection chamber (TPC) located at the Waste Isolation Plant Project (WIPP) in New Mexico, USA. The detector consists of two almost identical TPC halves with a shared, optically transparent, cathode \cite{Aug12a}, which creates two drift regions with a drift field of $\sim400$\,V/m. The detector is filled with 175\,kg of liquid xenon enriched to $\sim81\%$ in the isotope $^{136}$Xe. A sectioned view of an engineering rendering of EXO-200 is shown on the right side in Figure\,\ref{fig:EXO-200}. Radioactive decays and cosmic radiation deposit energy in the detector volume, ionizing the xenon and creating scintillation light and free electrons, which are drifted towards the anode wire planes. Both, scintillation light and electric charge are read by large-area avalanche photo diodes (APDs) and two wire planes, called $u$ and $v$ wires, respectively. Scintillation-light and charge measurement are used to fully reconstruct the energy of each event, its location within the detector volume, and its multiplicity, i.e., the number of locations at which energy was deposited in each event. Beta events deposit energy predominantly in on location (single-site events), while $\gamma$s scatter depositing energy at multiple locations (multi-site events). Figure\,\ref{fig:EXO-200} shows the single-site energy spectrum of EXO-200 which is dominated by $2\nu\beta\beta$ events. The multi-site spectrum (not shown) mainly consists of $\gamma$ events, which are used to constrain the background models of the single-site fit. Alpha events mainly emit scintillation light and are easily identified and discriminated. The event location allows to optimize the sensitivity of a physics search by adjusting the fiducial volume and taking advantage of the self shielding of xenon.

In phase I of data taking with EXO-200, an energy resolution of $1.53\pm0.06$\% at the $Q$-value was achieved \cite{Albert2014}. This data set allowed the measurement of the $2\nu\beta\beta$ half life $T^{2\nu}_{1/2}=2.165 \pm 0.061 \cdot 10^{21}$\,years \cite{Albert2014PRC}, which is the slowest decay rate ever measured directly, and put a limit on the $0\nu\beta\beta$ half life of $T^{0\nu}_{1/2}>1.1\cdot 10^{25}$\,years at the 90\%  confidence level (C.L.) with a sensitivity of $1.9\cdot10^{25}$\,years at an exposure of 100\,kg-years \cite{Albert2014}. 

In early 2014, two independent incidents at the WIPP site caused the mine operation and EXO-200 to halt, and only in early 2016 EXO-200 low-background data taking could resume. During this down-time part of the detector's front-end electronics was upgraded and a "deradonator" was installed to reduce the radon concentration in the air gap between the outer cryostat and low-radioactive lead shielding. These upgrades are expected to further improve EXO-200's sensitivity by a factor 2 to 3.
\subsection{KamLAND-Zen}
KamLAND is a monolithic liquid scintillator detector located in the Kamioka Mine in Japan. The original oscillation experiment used one kiloton of liquid scintillator contained in a 6.5~m radius balloon to detect antineutrinos from Japan's nuclear reactors. KamLAND-Zen uses this large scintillating volume as an active shield for a central volume of enriched xenon-doped liquid scintillator contained in an inner balloon with a radius of 3~m. 

The inner balloon was installed in 2011, the same year as the great east Japan earthquake and subsequent Fukushima power plant disaster. The first phase of data taking from October 12, 2011 to June 14, 2012 with an exposure of 89.5~kg-year of $^{136}$Xe showed a significant contamination from $^{110m}$Ag, a fission product. A purification campaign successfully removed this background. The post-purification phase collected data from December 11, 2013 to October 27, 2015, corresponding to an exposure of 504 kg-year and a sensitivity of $5.6\times10^{25}$~years. It set a limit of $T^{0\nu}_{1/2} > 9.2\times10^{25}$~years and when combined with the pre-purification data set leads to a limit of $T^{0\nu}_{1/2}>1.01\times 10^{26}$ at the 90\% C.L. This is the leading limit for $0\nu\beta\beta$ and KamLAND-Zen is the first experiment to surpass the $10^{26}$~year half-life.

The success of KamLAND-Zen has shown that the advantages of the liquid scintillator technique: large masses, self-shielding and fully contained energy depositions, can make up for the relatively poor energy resolution. The KamLAND-Zen experiment is in the process of installing a new slightly larger mini-balloon to hold $\sim$800~kg of $^{136}$Xe. Data taking for KamLAND-Zen 800 is expected to start in the next year. This is the first step in KamLAND-Zen's ton-scale program that includes a major upgrade to the detector described in the next section.
\begin{figure}
\centering
\includegraphics[trim={0mm 55mm 0mm 0mm}, width=.9\textwidth]{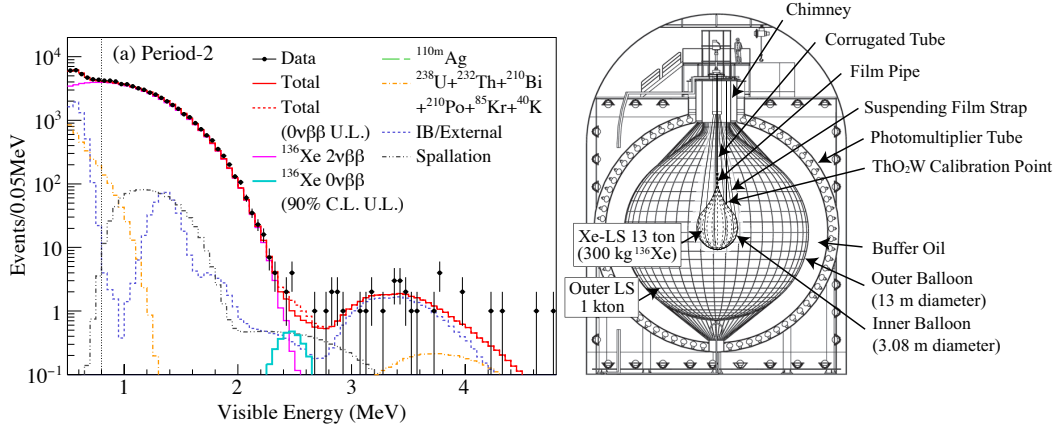}
\caption{(left) KamLAND-Zen Energy spectrum of selected $0\nu\beta\beta$ candidates within a 1-m-radius spherical volume in Period-2 drawn together with best-fit backgrounds, the $2\nu\beta\beta$ decay  spectrum,  and  the  90\%  C.L. upper limit for $0\nu\beta\beta$ decay from Ref.~\cite{KamLAND-Zen:2016pfg}. (right) Schematic diagram of the KamLAND-Zen detector from Ref.~\cite{KamLANDZen:2012aa}.}
\label{fig:KZ}
\end{figure}
\section{Tonne-Scale Experiments}
In order to make a definitive search for $0\nu\beta\beta$ in the inverted mass-hierarchy, target masses on the order of a few tonnes are required. Several collaborations are developing detector concepts to probe this parameter space. KamLAND2-Zen and nEXO propose to search for $0\nu\beta\beta$ in $^{136}$Xe deploying 1\,tonne and 5\,tonnes of enriched Xe, respectively. Their projected sensitivity limit is show as a cross-hatched band in Figure\,\ref{fig:sensitivity-comp}.
\subsection{nEXO}
\begin{figure}
	\centering
		\includegraphics[width=.5\textwidth]{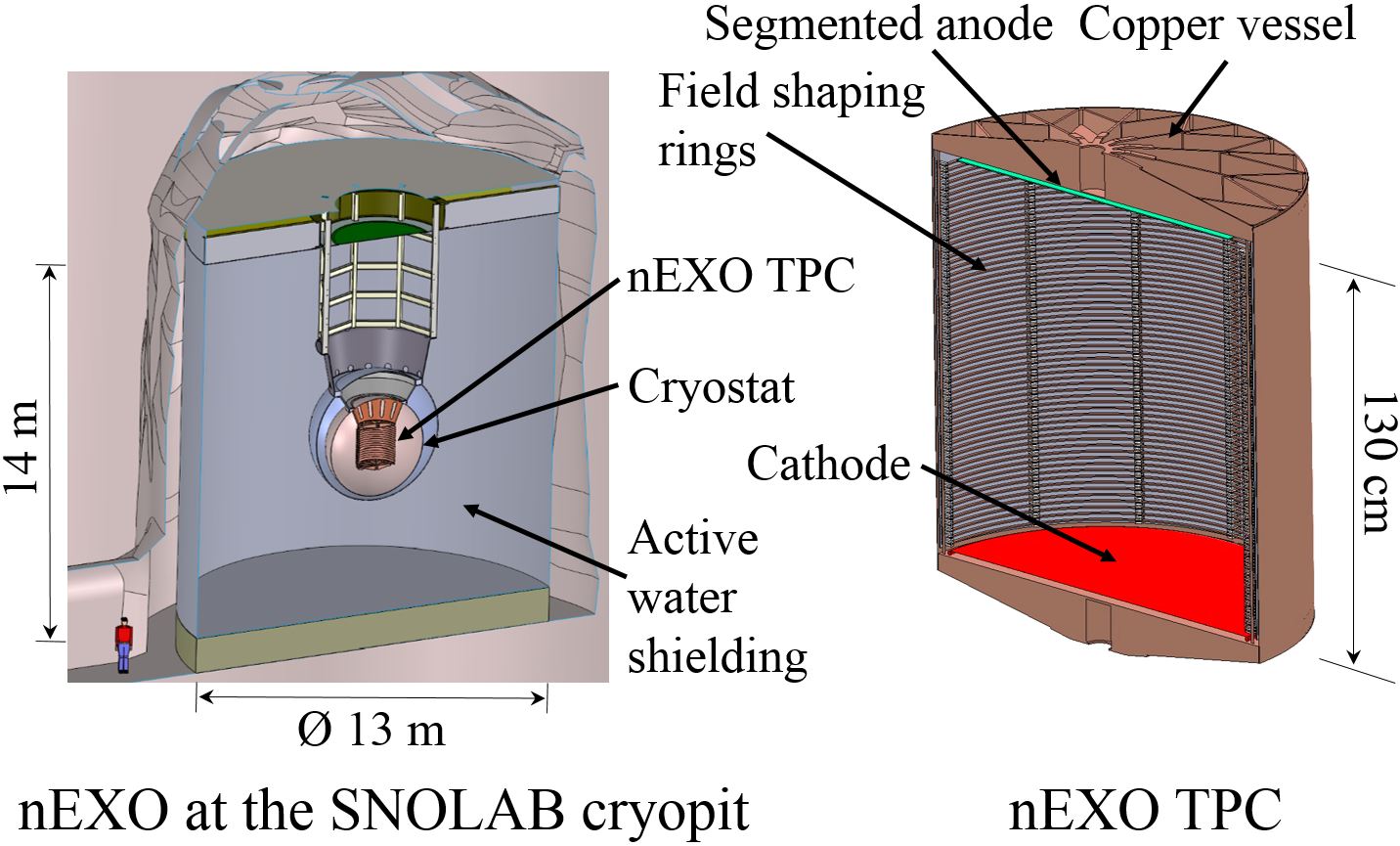}
	\caption{Artist rendering of the nEXO TPC (right) and its installation at the SNOLAB cryopit (left). The cryostat is submerged in a water tank which acts as active shielding.  SiPMs will be mounted between field shaping rings and detector wall.
	\label{fig:nexo}}
\end{figure}

The nEXO detector concept is based on the success of EXO-200. The detector is anticipated to be deployed at SNOLAB in Ontario, Canada, where the Nobel Prize winning SNO detector was located. An artist rendering of the detector is shown in Figure\,\ref{fig:nexo}. nEXO is being designed as a cylindrical, monolithic, single-volume, liquid xenon TPC with a drift field of 400\,V/cm deploying 5 tonnes of xenon enriched in $^{136}$Xe at $\sim$90\%. A segmented anode with perpendicular $x$ and $y$ channels collects the charge signal, while scintillation light is recorded by Silicon Photon Multipliers (SiPMs). These photon detectors are mounted outside of the field shaping rings, but inside the liquid xenon volume, covering the area of the cylindrical detector wall ($\sim4\,\textnormal{m}^2$ area). SiPM devices sensitive to Xe-scintillation light at 175\,nm are currently being developed and tested by the nEXO collaboration \cite{Ostrovskiy2015}. 

Simulations of nEXO, based on EXO-200 and radio-assay data, predict a sensitivity to $T_{1/2}^{0\nu}$ of $9.5\cdot 10^{27}$\,years after 10\,years of data taking. The resulting sensitivity to the effective Majorana neutrino mass is shown as a cross-hatched band in Figure\,\ref{fig:sensitivity-comp} for different matrix elements \cite{Rodriguez2010,Mustonen2013}. This assumes an improved energy resolution of 1\% at the $Q$-value, which is achieved by improved light detection and electronics. nEXO, like EXO-200, will fully reconstruct event energy, location, multiplicity, and topology. This sophisticated reconstruction in conjunction with the self-shielding of xenon will significantly increase nEXO's sensitivity in comparison to an experiment deploying a similar target mass of a different isotope in many small-volume detectors.
%
\subsection{KamLAND2-Zen}
KamLAND2-Zen's focus is an improvement in the energy resolution from $\sim4\%\sqrt{E}$ to $~\sim2\%\sqrt{E}$. The detector has been running continuously for more than 15 years and is due for a major refurbishment. As part of this work the detector will be drained and the main spherical tank will be inspected. The improvement in energy resolution comes from the installation of new high quantum efficiency photomultiplier tubes with Winston cones and higher light yield LAB-based liquid scintillator. The R\&D indicates that these three improvements boost the light collection efficiency by factors of 1.9, 1.8 and 1.4 respectively.

The improvement in energy resolution is complemented by a modest increase in the isotope mass to bring the total to 1~tonne and new electronics to improve the tagging of the muon spallation background from $^{10}$C. More novel background techniques involving scintillating balloon film and a secondary imaging system are also being explored. The goal of the KamLAND2-Zen phase is to reach 20~meV.
\section{Beyond the Ton-Scale}
The increase in sensitivity in future $0\nu\beta\beta$ searches will be limited by the available target mass. Advanced technologies may provide a path forward towards probing further into the normal neutrino-mass hierarchy. These technologies must suppress $\beta$, $\gamma$ and even solar neutrino backgrounds which ultimately limit a detector's sensitivity to $0\nu\beta\beta$. 

Two approaches are presented with the potential to identify $\beta\beta$ events by either probing the decay volume for the existence of the $^{136}$Xe decay daughter $^{136}$Ba, or by applying directionally sensitive liquid scintillator.
\subsection{Barium Tagging for $0\nu\beta\beta$ searches}
Ba-tagging is being developed for application in a monolithic xenon TPC and describes the following concept: when a $0\nu\beta\beta$-candidate event is recorded  
it is localized instantly and a small volume surrounding the event's location is extracted from the detector volume and probed for the presence of a Ba-ion. If a $^{136}$Ba is found, the event is considered for the $0\nu\beta\beta$ search, otherwise it is classified as background. This unambiguous identification of events at the $Q$-value as $\beta\beta$ or background events increases the detector's sensitivity without increasing its mass, however, more significant is the ability to confirm an observed $0\nu\beta\beta$ signal as originating from true $\beta\beta$-decay events.

\begin{figure}
	\centering
		\includegraphics[width=0.9\textwidth]{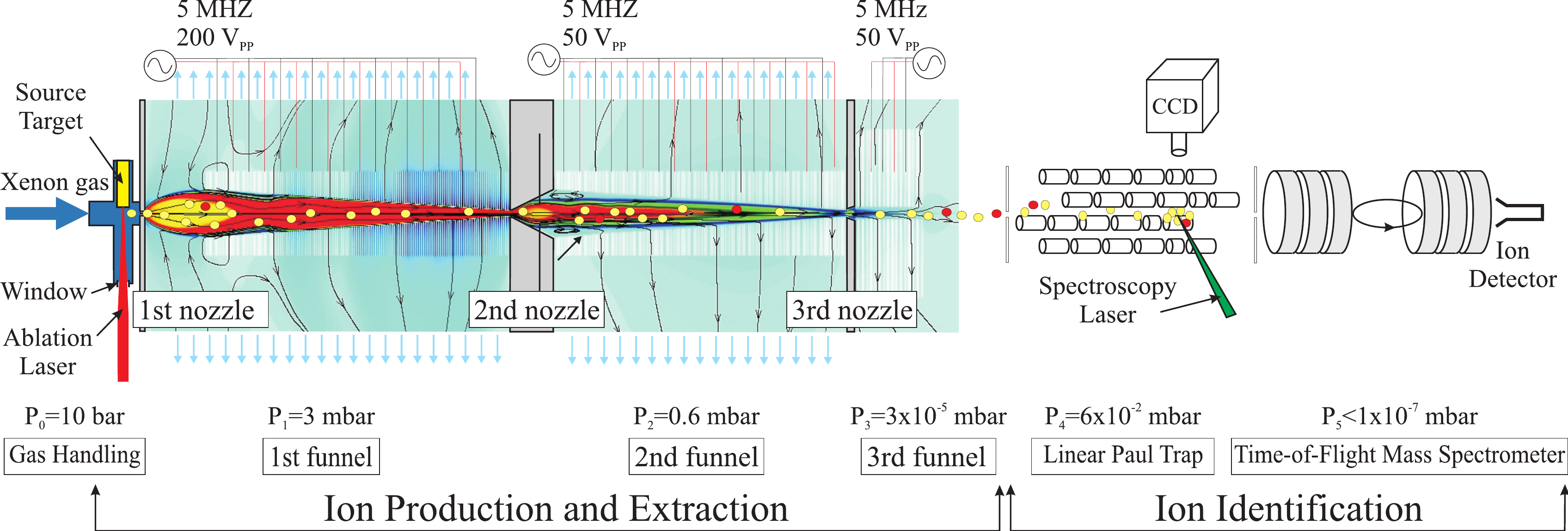}
	\caption{Schematic of the setup to extract Ba-ions from xenon gas and identify them by means of laser-fluorescence spectroscopy. Ba-ions are produced through laser-ablation at a source target located in xenon gas and extracted into vacuum by a combination of RF funnels. The mass spectrometer is proposed for systematic ion-extraction studies. Gas-flow calculation courtesy of V. Varentsov.}
	\label{fig:RF-funnel}
\end{figure}

Ba-tagging has been proposed by \cite{Moe1991} and various approaches are pursued by the nEXO collaboration using a tip or cold probe to extract $^{136}$Ba from the volume \cite{Twelker2014a,Mong2015} (see \cite{Jones2016} for a proposed technique to identify Ba inside the detector). An alternative approach proposes to move a capillary close to the event location and flush the $^{136}$Ba-ion out of the detector with liquid xenon. Once outside the detector, the xenon undergoes a phase transition and a radio-frequency (RF) ion funnel is applied to separate ions from the neutral xenon gas. Following the extraction into vacuum, the Ba-ion will be captured in a linear Paul trap and identified through isotope-selective laser-fluorescence spectroscopy. Such a system is currently being developed collaboratively by Carleton University, McGill University, and TRIUMF and is based on an RF ion funnel developed at Stanford University. This RF-funnel allowed the extraction of ions, produced by either a $^{148}$Gd $\alpha$ or a $^{252}$Cf fission source, from xenon gas of up to 10\,bar into vacuum \cite{Brunner2015}, achieving for the first time ion extraction from such high pressures. In parallel, an element-sensitive fluorescent laser spectroscopy technique on Ba-ions trapped in a linear Paul trap has been developed and individual trapped Ba-ions were identified \cite{Green2007}. The RF-funnel ion-extraction and fluorescent laser spectroscopy setup will be further improved and combined to demonstrate the feasibility of the proposed approach. For future studies, the radioactive ion source will be replaced with a surface laser-ablation ion source to selectively create Ba ions for extraction studies. A schematic of the proposed setup is shown in Figure\,\ref{fig:RF-funnel}. A multi-reflection time-of-flight mass spectrometer will be added to allow for identification of ions other than barium, which is of interest for ion-extraction and ion-transport studies. 

Significant progress has been made in the development of Ba-tagging techniques, especially Ba-ion identification through laser-fluorescence spectroscopy has achieved a sensitivity on the one to three ion level. 
Further efforts will focus on implementing this development in a workable Ba-tagging technique, which is sensitive to individual ions in tonnes of xenon. Ba-tagging is a great challenge, however, the significant increase in sensitivity to $T_{1/2}^{0\nu}$ and the possibility to identify events as true $\beta\beta$ decays makes it a powerful tool that can be applied to a future nEXO-like detector.
\subsection{Directional Liquid Scintillator}
The next step in the KamLAND program is Super-KamLAND-Zen with a target sensitivity of 8~meV. The increased sensitivity comes from installing 40~tonnes of pressurized xenon-doped liquid scintillator into the center of Super-Kamiokande. This would be coordinated with the start of Hyper-Kamiokande. The pressurization is needed to increase the xenon concentration in the mini-balloon. This allows an increase in mass while maintaining background levels since most backgrounds scale with volume. The R\&D has already begun both on the light yield of this pressurized xenon scintillator and on the engineering of this balloon.

At this size and target sensitivity, the background from $^8$B solar neutrinos is non-negligible. The ability to reconstruct the direction of the $\sim$MeV electrons would be a powerful background rejection tool for both this and other backgrounds and would be revolutionary for large-scale liquid scintillator detectors in general.  

\begin{figure}
	\centering
		\includegraphics[trim={0mm 60mm 0mm 0mm 0mm},clip, width=.9\textwidth]{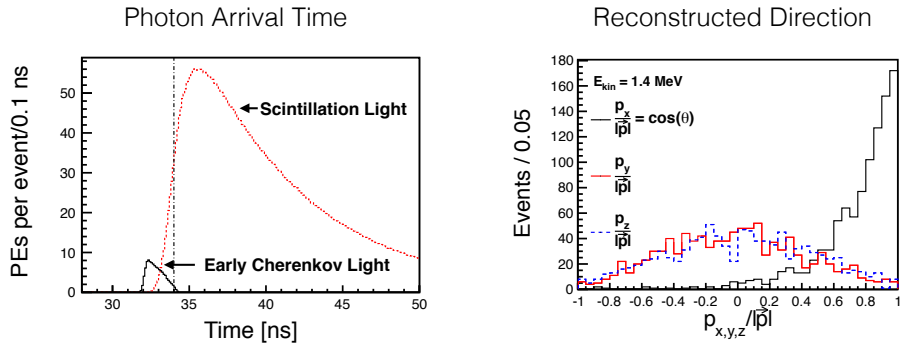}
\caption{(left)Photoelectron (PE) arrival times after application for the simulation of 1000 electrons (5 MeV). PEs from Cherenkov light (black, solid line) and scintillation light (red, dotted line) are compared. The dash-dotted vertical line illustrates a time cut at 34.0 ns. This is the default simulation: bialkali photocathode and TTS = 0.1 ns $(\sigma)$.  After the 34.0 ns time cut, 171 PEs from scintillation and 108 PEs from Cherenkov light are detected. (right)  The reconstructed direction, $(p_x/|\vec{p}|, p_y/|\vec{p}|, p_z/|\vec{p}|)$, for the simulation of 1000 electrons. In the simulation the electrons are produced along the x-axis, $\vec{p}/|\vec{p}|$ = (1,0,0), and originate at the center of the 6.5m-radius detector, $\vec{r}$ = (0,0,0). Only photons with arrival time of $t<$ 34.0~ns are used in the reconstruction. The quantum efficiency of the bialkali photocathode is taken into account. The reconstruction at 1.4 MeV is shown. From Ref.~\cite{Aberle:2013jba}.}
	\label{fig:NuDot}
\end{figure}

Direction reconstruction relies on the ability to separate the Cherenkov and scintillation light. The composition of a liquid scintillator cocktail determines an absorption cutoff, photons with wavelengths' shorter than this wavelength become part of the isotropic scintillation light, but photons with wavelengths longer than this cutoff propagate undisturbed and retain their directional information. Ref.~\cite{Aberle:2013jba} showed that this separation could be obtained and the direction of $\sim$MeV electrons could be reconstructed if photo detectors with $\sim$100~ps timing were used, see Fig.~\ref{fig:NuDot}. The separation is improved with red-sensitive photo cathodes and scintillator emission spectra narrowed using novel wavelength shifters like quantum-dots. The CHESS experiment recently demonstrated Cherenkov/Scintillation separation in an LAB-based cocktail using a bench-top apparatus and cosmic muons \cite{Caravaca:2016fjg}. A prototype detector called NuDot is being constructed at MIT to demonstrate this technique on the ton-scale.
\section{Conclusion}
The search for $0\nu\beta\beta$ decays is an exciting quest to investigate if neutrinos are Majorana particles. Current generation experiments, such as KamLAND-ZEN and EXO-200, are probing the degenerate hierarchy parameter space down to about 60\,meV. With the advent of tonne-scale experiments, this limit will be pushed below 10\,meV, hence fully probing the inverted mass hierarchy. Depending on the nature of the neutrino, a ground-breaking discovery is within reach of next generation $0\nu\beta\beta$ detectors. It is an exciting time for $0\nu\beta\beta$ decay searches.
\section{Acknowledgment}
T.B. and L.W. acknowledge support from the EXO-200 \& nEXO, and KamLAND-ZEN collaborations, respectively.
%
%
\begin{spacing}{0.9}
\bibliographystyle{ieeetr}
\bibliography{sample}

\begin{thebibliography}{10}

\bibitem{Tretyak:2002dx}
V.~I. Tretyak and Y.~G. Zdesenko, ``{Tables of double beta decay data: An
  update},'' {\em Atom. Data Nucl. Data Tabl.}, vol.~80, p.~83, 2002.

\bibitem{DellOro2016}
S.~Dell’Oro, S.~Marcocci, M.~Viel, and F.~Vissani, ``{Neutrinoless Double
  Beta Decay: 2015 Review},'' {\em Advances in High Energy Physics}, vol.~2016,
  p.~2162659, 2016.

\bibitem{Mustonen2013}
M.~T. Mustonen and J.~Engel, ``{Large-scale calculations of the
  double-$\ensuremath{\beta}$ decay of $^{76}$Ge, $^{130}$Te, $^{136}$Xe, and
  $^{150}$Nd in the deformed self-consistent Skyrme quasiparticle random-phase
  approximation},'' {\em Physical Review C}, vol.~87, p.~064302, 2013.

\bibitem{Rodriguez2010}
T.~R. Rodr\'{\i}guez and G.~Mart\'{\i}nez-Pinedo, ``{Energy Density Functional
  Study of Nuclear Matrix Elements for Neutrinoless
  $\ensuremath{\beta}\ensuremath{\beta}$ Decay},'' {\em Physical Review
  Letters}, vol.~105, p.~252503, 2010.

\bibitem{Albert2014}
{J. B. Albert et al.}, ``{Search for Majorana neutrinos with the first two
  years of EXO-200 data},'' {\em Nature}, vol.~510, p.~229, 2014.

\bibitem{Albert2014PRC}
{J.B. Albert et al.}, ``{Improved measurement of the $2\nu\beta\beta$ half-life
  of $^{136}$Xe with the EXO-200 detector},'' {\em Physical Review C}, vol.~89,
  p.~015502, 2014.

\bibitem{Aug12a}
{M. Auger et al.}, ``{The EXO-200 detector, part I: detector design and
  construction},'' {\em Journal of Instrumentation}, vol.~7, p.~P05010, 2012.

\bibitem{KamLAND-Zen:2016pfg}
A.~Gando {\em et~al.}, ``{Search for Majorana Neutrinos near the Inverted Mass
  Hierarchy Region with KamLAND-Zen},'' {\em Physical Review Letters},
  vol.~117, p.~082503, 2016.
\newblock [Addendum: Phys. Rev. Lett. 117, no.10, 109903(2016)].

\bibitem{KamLANDZen:2012aa}
A.~Gando {\em et~al.}, ``{Measurement of the double-$\beta$ decay half-life of
  $^{136}$Xe with the KamLAND-Zen experiment},'' {\em Physical Review C},
  vol.~C85, p.~045504, 2012.

\bibitem{Ostrovskiy2015}
{I. Ostrovskiy et al.}, ``{Characterization of Silicon Photomultipliers for
  nEXO},'' {\em IEEE Transactions on Nuclear Science}, vol.~62, p.~1825, 2015.

\bibitem{Moe1991}
M.~K. Moe, ``Detection of neutrinoless double-beta decay,'' {\em Phys. Rev. C},
  vol.~44, p.~R931, 1991.

\bibitem{Twelker2014a}
{K. Twelker et al.}, ``{An apparatus to manipulate and identify individual Ba
  ions from bulk liquid Xe},'' {\em Review of Scientific Instruments}, vol.~85,
  p.~095114, 2014.

\bibitem{Mong2015}
{B. Mong et al.}, ``{Spectroscopy of Ba and ${\mathrm{Ba}}^{+}$ deposits in
  solid xenon for barium tagging in nEXO},'' {\em Phys. Rev. A}, vol.~91,
  p.~022505, 2015.

\bibitem{Jones2016}
B.~Jones, A.~McDonald, and D.~Nygren, ``Single molecule fluorescence imaging as
  a technique for barium tagging in neutrinoless double beta decay,'' {\em
  Journal of Instrumentation}, vol.~11, p.~P12011, 2016.

\bibitem{Brunner2015}
{T. Brunner et al.}, ``{An RF-only ion-funnel for extraction from high-pressure
  gases},'' {\em International Journal of Mass Spectrometry}, vol.~379, p.~110,
  2015.

\bibitem{Green2007}
{M. Green et al.}, ``{Observation of single collisionally cooled trapped ions
  in a buffer gas},'' {\em Phys. Rev. A}, vol.~76, p.~023404, 2007.

\bibitem{Aberle:2013jba}
C.~Aberle, A.~Elagin, H.~J. Frisch, M.~Wetstein, and L.~Winslow, ``{Measuring
  Directionality in Double-Beta Decay and Neutrino Interactions with
  Kiloton-Scale Scintillation Detectors},'' {\em Journal of Instrumentation},
  vol.~9, p.~P06012, 2014.

\bibitem{Caravaca:2016fjg}
J.~Caravaca, F.~B. Descamps, B.~J. Land, M.~Yeh, and G.~D. Orebi~Gann,
  ``{Cherenkov and Scintillation Light Separation in Organic Liquid
  Scintillators},'' {\em arXiv:1610.02011}, 2016.

\end{thebibliography}
\end{spacing}
\end{document}